\documentclass[twocolumn,prl,superscriptaddress,showpacs]{revtex4}
\usepackage{graphicx}% Include figure files
\usepackage{color}
\usepackage{ulem}
\begin{document}
\title{Dynamics of a Heisenberg spin chain in the quantum critical regime:
NMR experiment versus effective field theory}

\author{H.~K\"{u}hne}
\affiliation{Institut f\"{u}r Festk\"{o}rperphysik, TU Dresden,
01069 Dresden, Germany}
\author{A.A.~Zvyagin}
\affiliation{Institut f\"{u}r Festk\"{o}rperphysik, TU Dresden,
01069 Dresden, Germany} \affiliation{Institute for Low Temperature
Physics and Engineering of the NAS of Ukraine, Kharkov, 61103,
Ukraine}
\author{M.~G\"{u}nther}
\affiliation{Institut f\"{u}r Festk\"{o}rperphysik, TU Dresden,
01069 Dresden, Germany}
\author{A.P.~Reyes}
\affiliation{National High Magnetic Field Laboratory, Tallahassee,
Florida 32310, USA}
\author{P.L.~Kuhns}
\affiliation{National High Magnetic Field Laboratory, Tallahassee,
Florida 32310, USA}
\author{M.M.~Turnbull}
\affiliation{Carlson School of Chemistry and Department of Physics,
Clark University, Worcester, Massachusetts 01610, USA}
\author{C.P.~Landee}
\affiliation{Carlson School of Chemistry and Department of Physics,
Clark University, Worcester, Massachusetts 01610, USA}
\author{H.-H.~Klauss}
\affiliation{Institut f\"{u}r Festk\"{o}rperphysik, TU Dresden,
01069 Dresden, Germany}

\date{\today}

\begin{abstract}
A comprehensive comparison between the magnetic field- and
temperature-dependent low frequency spin dynamics in the
antiferromagnetic spin-1/2 Heisenberg chain (AFHC) system copper
pyrazine dinitrate, probed via the $^{13}$C-nuclear magnetic
resonance (NMR) relaxation rate $T_1^{-1}$, and the field
theoretical approach in the Luttinger liquid (LL) regime has been
performed. We have found a very good agreement between the
experiment and theory in the investigated temperature and field
range. Our results demonstrate how strongly the quantum critical
point affects the spin dynamics of Heisenberg spin chain
compounds.
\end{abstract}

%\keywords{}

\maketitle

The occurrence of quantum phase transitions (QPT) in systems of
correlated electrons is a very important topic in current
solid-state physics. These transitions are present in, e.g.,
high-T$_c$ superconductors, heavy-fermion metals, or magnetic
insulators \cite{NatureFocus2008}. The phase diagrams of systems
from the first two classes are in general complex due to several
interaction mechanisms. In contrast, the purely magnetic
interactions in magnetic insulators, in particular for
one-dimensional spin systems, give the rare occasion to perform
exact calculations of their characteristics and to compare them with
experimental data sets of well characterized sample systems
\cite{Klanjsek}. The spin properties of organic-based
low-dimensional magnets can be fine-tuned by chemical synthesization
methods. This well controlled synthesis allows the systematic
investigation of the magnetic properties with well established
methods such as neutron scattering, ESR, dc/ac magnetometry, $\mu$SR
or NMR. In low-dimensional magnets the high sensitivity of NMR
\cite{Borsa} to local hyperfine fields allows to perform detailed
studies of, e.g., phase transitions, or the local distribution of
spin moments.

The isotropic AFHC model is one of the main paradigms of quantum
many-body physics both from the experimental and theoretical
viewpoint. Its static characteristics were successfully compared
with experimentally studied features of quasi-one-dimensional
magnetic compounds, synthesized recently \cite{book}. For
dynamical properties, especially in the vicinity of the quantum
critical point, there is still an insufficiency of experimental
data sets for the low frequency part and their comparison to
calculations.

In this Letter we present a detailed comparison of field theory
results with the data of recent NMR experiments \cite{Kuehne}
probing the spin dynamics in a wide field- and temperature range in
one of the best realizations of the AFHC model, namely
Cu(C$_4$H$_4$N$_2$)(NO$_3$)$_2$ (also known as copper pyrazine
dinitrate or CuPzN) \cite{lancaster}. We find an extraordinary good
agreement in the low temperature behavior across the field driven
QPT.

Due to a relatively low value of the coupling constant
$J/k_B$=10.7 K in CuPzN, the critical field value $B_s = 14.6$ T
is well reachable by standard laboratory equipment. Therefore, one
can examine the spin dynamics in the region of fields and
temperatures, where spin-spin correlations manifest themselves in
the most prominent way, and compare with the results of various
theoretical methods. The inter-chain interactions are supposed to
be small, so that the magnetic ordering ($T_c\sim 107$~mK) did not
affect the AFHC behavior down to the lowest $T$ studied in the
NMR-experiment. Whereas these data were compared with numerical
Quantum Monte Carlo (QMC) simulations, their agreement with
results of a field theory approach (which serves as a very good
description namely at low $T$, where QMC simulations often produce
larger errors) for the LL regime is checked in our work. A similar
approach has been used to calculate the properties of several
low-dimensional systems \cite{Klanjsek}, but so far no comparison
with NMR-data of a direct realization of the AFHC has been done,
especially in the vicinity of the QPT.

The NMR relaxation rate $T^{-1}_1$ can be presented as
\cite{Moriya1956} $T_1^{-1} = (\gamma_e^2\gamma_N^2\hbar^2/2) \int
dq [F^{z}(q) S^{zz}(q,\omega_N) +F^{x}(q)S^{xx}(q,\omega_N)]$,
where $\gamma_{e}$ and $\gamma_{N}$ are the electronic and nuclear
gyromagnetic ratios, respectively, $\omega_N$ is the resonance
frequency of nuclear spins, $F^{z,x}(q)$ are the hyperfine
form-factors of nuclear spins, parallel and perpendicular to the
external dc magnetic field $B$, $S^{\mu\nu}(q,\omega_N)$ ($\mu\nu
= z,x$) are the components of the tensor of the dynamical
structure factor (DSF) of the AFHC, also parallel and
perpendicular to $B$. For the transverse components we have
$S^{xx}=S^{yy}$, because of the rotational symmetry perpendicular
to the field direction. Since $\omega_N \ll J/\hbar, \gamma_e B$,
we use the limit $\omega_N \to 0$. The ground state behavior of
the DSF of the AFHC in the external magnetic field has been
studied in \cite{MTBB}, where it was supposed that continua of
low-energy gapless excitations of the AFHC (known as spinons)
yield the main contribution to the DSF. That conjecture, supported
by numerical calculations for small-size spin chains \cite{MTBB},
was later confirmed by the exact calculation, which used Bethe
ansatz equations in the ground state of the AFHC \cite{CM}: About
75\% of the DSF of the AFHC is determined by the two-spinon
continuum. At $B=0$ the DSF is isotropic, and at $\omega \to 0$
two points, $q=0$ and $q=\pi$, contribute mostly to the relaxation
rate. $B \ne 0$ introduces anisotropy in the components of the DSF
of the AFHC. For $B\ne 0$ the main contribution to the components
of the DSF of the spin chain is determined by the lower and upper
boundaries of the two-spinon continua \cite{MTBB}. The nonzero
field causes the shift of the most important contribution for
$S^{zz}(q,\omega=0)$ from $q = \pi$ to $q = \pi(1-2m)$, where $m$
is the magnetic moment per site of the AFHC, while the point $q=0$
remains important. For $S^{xx}(q,\omega=0)$ the point $q=\pi$
remains relevant, while instead of $q=0$ the main contribution
comes from $q=2\pi m$.

The asymptotic behavior of the correlation functions of the AFHC can
be calculated in the framework of the conformal field theory
\cite{Zb}. The low-energy states approach zero ($\omega=0$) at the
vector $q\sim P_F$, where $P_F$ is the Fermi momentum. For the
longitudinal component of the DSF at $q \sim P_F=0$ we have
\begin{equation}
\pi v^2 S_h^{zz} = 2K \alpha k_BT \ , \label{Szz0}
\end{equation}
where $K$ is the LL exponent, $v$ is the Fermi velocity of a
spinon, and $\alpha$ is the cut-off parameter of the theory. For
$q \sim P_F =\pi(1-2m)$ we get
\begin{equation}
{v S_s^{zz}\over \cos(2\pi K)} \sim C_1B^2\left( {K\over 2}, 1-K \right) \left({2\pi
\alpha k_BT\over
v}\right)^{2K-1}  , \label{Szzpi}
\end{equation}
where $B(x,y)=\Gamma(x)\Gamma(y)/\Gamma(x+y)$ is the beta
function, and $C_1$ is the field-dependent multiplier \cite{LZ}.
At $m=1/2$ this contribution has to coincide with
Eq.~(\ref{Szz0}), which defines $C_1(B_s)$. The calculated
longitudinal components of the DSF manifest weak dependencies on
$T$ and $B$, except in the vicinity of the QPT (at which $v \to
0$), where they show a strong growth linear in $T$.

For the transverse component of the DSF at $q \sim P_F=\pi$, we have
\begin{equation}
{v S_s^{xx} \over |\cos(2\pi \gamma')|} \sim C_2
B^2\left( {\gamma'\over 2}, 1-\gamma' \right)\left({2\pi \alpha k_BT\over
v}\right)^{2\gamma'-1} ,
\label{Sxxpi}
\end{equation}
where $\gamma' = 1/(4K)$. Near the saturation point $B=B_s$ the
correlation amplitude goes to zero, while the value of the
correlation function at zero field in the ground state is
approximately equal to 0.18 \cite{LZ}. Hence, we can write the
multiplier as $C_2 = 0.18/B^2(1/4,1/2) \approx 0.0065$. This
component yields the main contribution to the measured relaxation
rate, see below, and the results of its calculation are presented in
Fig.~1.
\begin{figure}
\begin{center}
\includegraphics[width=0.7\columnwidth]{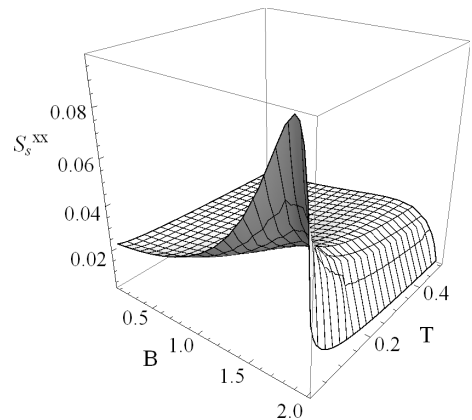}
\end{center}
\caption[1]{The magnetic field (given in units of $J
/\hbar\gamma_e$) and temperature (given in units of $k_B/J$)
behavior of the transverse staggered component of the DSF of the
AFHC model, calculated using the conformal field theory. }
\label{fig1}
\end{figure}
For $q \sim P_F=2\pi m$ we get
\begin{eqnarray}
&&{v S_h^{xx}\over \cos(2\pi\gamma)} \sim  2C_3
B\left( {\gamma +2\over 2}, -1-\gamma \right)B\left(
{\gamma\over 2}, 1-\gamma \right)\nonumber \\
&&\times \left({2\pi \alpha k_BT\over v}\right)^{2\gamma+1}
 \ , \label{Sxx0}
\end{eqnarray}
where $\gamma = K -1 +1/(4K)$. In the ground state numerical
calculations \cite{LZ} give the value of the correlation function at
zero field $\sim 0.03$, which defines $C_3(B=0)$. At $B=0$ the
exponent for the AFHC is $K=1/2$, \cite{Zb} and, therefore, this
component of the DSF, calculated in this approach, diverges at
$B=m=P_F=0$. However, that divergency is well-known to be
nonphysical. It is easy to calculate the transverse homogeneous
magnetic susceptibility for the Heisenberg spin system at $q=0$; it
is equal to $m/\hbar \gamma_e B$. For the AFHC the magnetic moment
is proportional to the field for small values of $B$, hence, in that
region the transverse magnetic susceptibility coincides with the
longitudinal one. The magnetic susceptibility is related to the DSF
via the fluctuation-dissipation theorem. The correct magnetic field
behavior of this transverse component of the DSF has to behave as
the homogeneous longitudinal component for small values of the
field, and decay to zero at $B \to B_s$, i.e. to coincide with the
staggered transverse component there, because at $B=B_s$ we have
$m=1/2$, and $P_F=2\pi m =\pi$.

The velocity and the LL exponent of spinons are $B$-dependent.
These dependencies can be obtained from the exact Bethe ansatz
solution. Recently, a simple ansatz for the magnetic field
behavior of the velocity $v$ and exponent $K$, valid in the
interval $0 \le B \le B_s$, was proposed \cite{Z}: $v = (\pi J/2)
\sqrt{[1-(B/B_s)][1-(B/B_s)+(2\hbar \gamma_e B/\pi J)]}$, $K =
f/\sqrt{4f^2 - 3(\hbar \gamma_e B)^2}$, $f=\pi J[1 - (B/B_s)] +
\hbar \gamma_eB$. The behavior of $v$ and $K$, given by those
expressions, agrees with the Bethe ansatz calculations. Finally,
marginal operators (from the renormalization group viewpoint)
introduce logarithmic corrections to the asymptotic behavior of
correlation functions of the AFHC in the conformal limit at low
$T$ \cite{Zb}. Those corrections can be taken into account, see,
e.g., \cite{BETG}, which yields the additional multiplier
$\sqrt{\ln (24.27J/ \alpha k_BT)}/(2\pi)^{3/2}$ to the right hand
sides of Eqs.~(\ref{Szz0})-(\ref{Sxx0}).

During the last years a new approach for the calculation of the
critical exponents was developed \cite{PKKG}. It was pointed out
that the low-energy dynamics of quantum chains is determined not
only by Fermi points, but also by high-energy states of the
system, i.e. the nonlinearity of the dispersion relations was
taken into account. Contributions from those high-energy states
were approximated as an interaction of the LL with an effective
impurity, which parameters are determined by quasi-momenta of
excitations. The theory, which determines the renormalization of
exponents of quantum chains in the presence of impurities, was
presented in \cite{Zb}. Generalizing the approach of \cite{PKKG},
we conjecture that the $B$- and $T$-behavior of the DSF is
determined by Eqs.~(\ref{Szz0})-(\ref{Sxxpi}) with exponents,
renormalized due to high-energy excitations. In the conformal
field theory we replace $\Delta M \to (\Delta M -n_{imp})$,
$\Delta D \to (\Delta D -d_{imp})$, where $\Delta M$ and $\Delta
D$ are integers, determining the finite-size spectra of the chain,
and $n_{imp}$ and $d_{imp}$ are the parameters of that effective
impurity. The latter are defined as \cite{PKKG} $n_{imp} = \pm
(\sqrt{K} -1)$, and $d_{imp} =- (1/2\sqrt{K})n_{imp}$. The plus
sign corresponds to the negative ``valence'' of the impurity, i.e.
to the hole in the Fermi sea, and the minus sign corresponds to
the positive ``valence'', i.e. to the excitation above the Fermi
sea. The behavior of the longitudinal homogeneous component of the
DSF, Eq.~(\ref{Szz0}), is, obviously, not renormalized. For the
longitudinal staggered-like component of the DSF, in
Eq.~(\ref{Szzpi}) we need to replace the exponent $K \to (1/2)
-(1/2\sqrt{K}) +(1/4K) +(9K/16)-(3\sqrt{K}/4)$ for a high-energy
hole, and $K \to (1/2) -(1/2\sqrt{K}) +(1/4K) +(K/4)-(\sqrt{K}/2)$
for a high-energy excitation. In Eq.~(\ref{Sxxpi}) for the
transverse staggered component of the DSF, we need to replace
$\gamma' \to (1/2) -(\sqrt{K}/2) +(K/4)$ for a high-energy hole,
and $\gamma' \to (1/2) +(1/K) -(1/2\sqrt{K}) +(K/4) -(\sqrt{K}/2)$
for a high-energy excitation. Finally, for the transverse
``homogeneous'' component, in Eq.~(\ref{Sxx0}) we need to replace
the exponent $\gamma \to (9K/16) -(1/2) -(3\sqrt{K}/4)$ for a
high-energy hole, and $\gamma \to (1/K) +K -(1/2) +(\sqrt{K}/2) -
(1/2\sqrt{K})$ for a high-energy excitation. However, the $T$- and
$B$-dependencies of the components of the DSF, obtained within our
conjecture (taking into account the high-energy excitation/hole of
the AFHC), do not agree with the experimentally observed data.
This can be explained as follows. For the calculation of the NMR
relaxation rate one performs the integration with respect to
quasi-momenta, where only a small interval near $\omega=0$
contributes mostly. The behavior of exponents in those intervals
is not homogeneous: One expects the ``traditional'' LL exponent
near the Fermi point and renormalized exponents close to the edges
of the interval, with a smooth crossover between the exponents.
The calculation of that crossover is a subtle point, not yet
performed analytically. Thus, NMR experiments suggest, that in the
interval of integration the main contribution comes from the
region, where the ``traditional'' exponent is applicable.

The NMR measurements were performed in two different standard NMR
setups, each with a superconducting magnet, a $^4$He temperature
insert and a commercial/homebuild spectrometer. An
inversion-recovery pulse sequence was used to measure the
$^{13}$C-nuclear relaxation rate $T_1^{-1}$.
\begin{figure}
\begin{center}
\includegraphics[width=0.9\columnwidth]{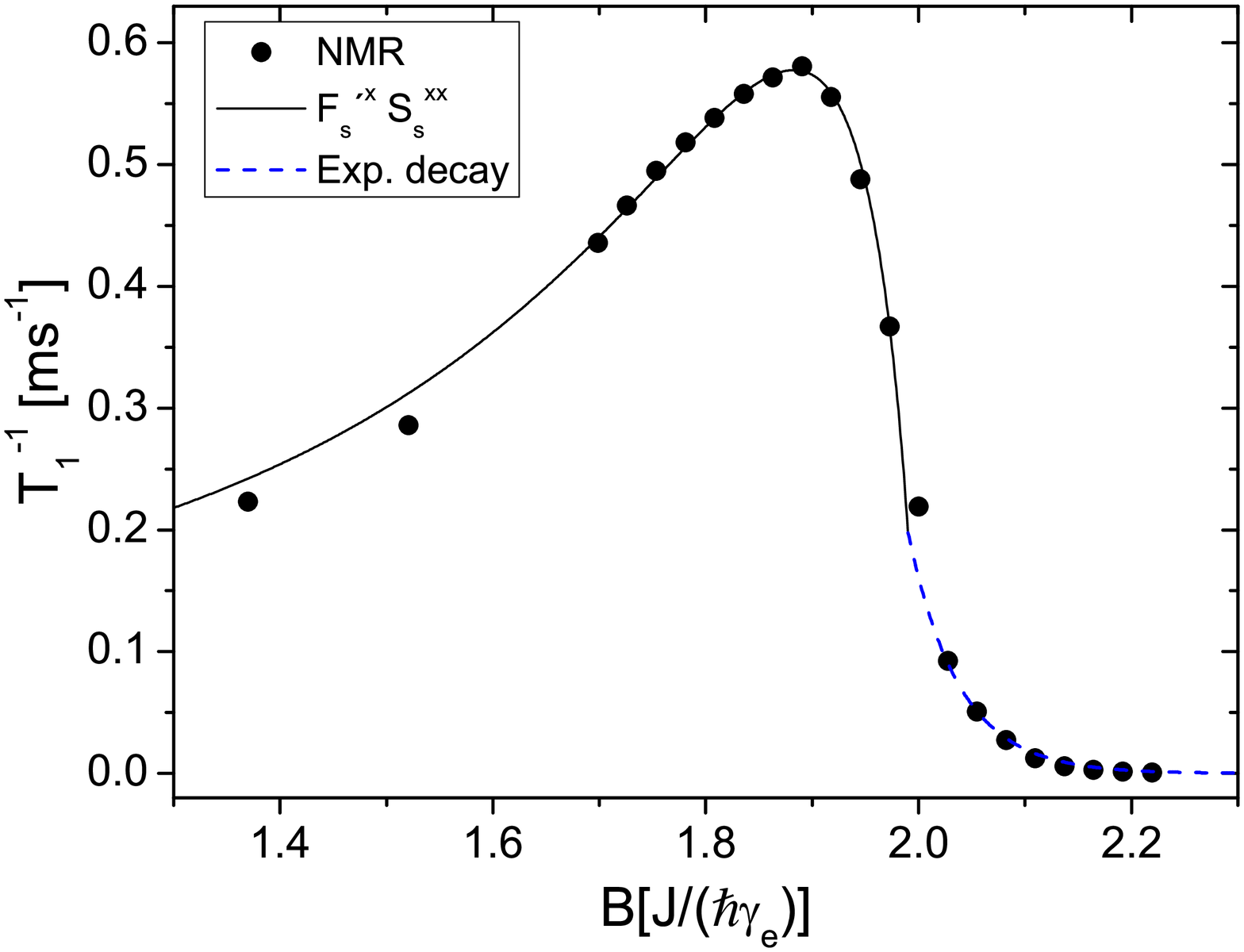}
\end{center}
\caption[1~]{ (Color online) Field dependence of ${T_1}^{-1}$ at
$k_B T / J =0.15$ (error bars are within symbol size). The results
of the effective field theory are given by the solid line for fields
up to $B_s$. Above $B_s$ an exponential function was fitted to the
data (dashed blue line). } \label{FieldDep14p6T}
\end{figure}
In order to ensure a well defined comparability with theory, the
hyperfine form-factor $F^z$, which relates $T_1^{-1}$ to $S^{zz}$,
was minimized. This was done by an orientation-dependent study of
the NMR frequency shift and $T_1^{-1}$, determining the angle of
$\angle(B,b)= 50^\circ$ in the $b-c$ plane \cite{Kuehne}. In this
orientation the critical field is $B_s = (2J/\hbar \gamma_e) =
14.6$ T, which is a slight adjustment to the previous value,
considering the anisotropic $g$-factor from recently published
ESR-results \cite{Validov2010}. In this case we have $T_1^{-1} =
F_s^{'x}(q)S_s^{xx} + F_h^{'x}(q)S_h^{xx}$, which leaves only two
form factors and the cut-off $\alpha$ as free parameters to fit
our calculations to the experimental data. The results of this fit
procedure are shown in Figs.~\ref{FieldDep14p6T} and \ref{fig2}.
As for the $B$-dependence shown in Fig.~\ref{FieldDep14p6T}, the
main contribution to the ${T_1}^{-1}$ rate comes from $S_s^{xx}$.
For fields larger than $B_s$, a spin excitation gap opens linear
with $B-B_s$, leading to an exponential decay of the relaxation
rate. We find an excellent agreement between our calculations and
the NMR experiment for the whole region of fields, in particular
near the QPT.

\begin{figure}
\begin{center}
\includegraphics[width=0.9\columnwidth]{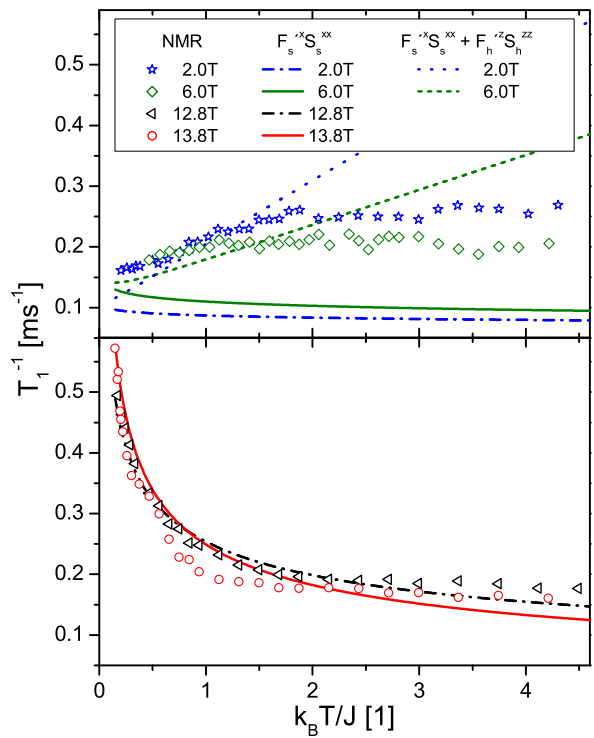}
\end{center}
\caption[1]{ (Color online) Comparison of temperature-dependent
experimental and theoretical ${T_1}^{-1} $-rates at different
external fields. At low fields, an additional contribution from the
homogeneous DSF is included in the theoretical rates, see text.}
\label{fig2}
\end{figure}
The temperature dependence of $T_1^{-1}$ at different fields is
shown in Fig.~\ref{fig2}. The values of $F_s^{'x}(q)$ and $\alpha$
obtained from the fit of the field dependence were used for the
scaling of $S_s^{xx}$ at all $B$ and $T$. At $B$ close to $B_s$,
the experimental and theoretical NMR-relaxation rates show a
diverging behavior as $T \to 0$. This singularity occurs when $B$,
acting as the chemical potential for spinons, crosses the boundary
of the dispersion relation (at this QPT $v$ goes to zero), cf.
Eq.~(\ref{Sxxpi}). Again, the critical regime, i.e. $T_1^{-1}(T)$
at 12.8T and 13.8T, can be fully described by
$F_s^{'x}(q)S_s^{xx}$. Fitting the experimental rates with a
power-law decay for $T < 2 J /\hbar \gamma_e$, in the region $0.7J
\le T \le 2J$ we find the exponents to be slightly smaller than
the theoretically calculated values , cf. Fig.~\ref{fig2}. Notice,
that one expects the accuracy of field-theoretical calculations to
significantly decrease for $T > J/k_B$. The agreement with
experiment is, nevertheless, surprisingly good for the whole
region of temperatures measured, with $T_1^{-1}$ becoming almost
$T$-independent at high temperatures. At low fields, the
experimental rates show an approximately linear $T$-dependence up
to $T \sim 2 J/k_B$. This behavior is not reflected by
$F_s^{'x}(q) S_s^{xx}$. As described above, for low fields
$S_h^{xx}$ has to be equivalent to $S_h^{zz}$. The fit of ${1/
T_1} = F_s^{'x}(q)S_s^{xx} + F_h^{'z}(q)S_h^{zz}$ for T$< 2 J /
\hbar \gamma_e$, with $F_h^{'z}$ as a free parameter gives much
better agreement with the experimental data in this low-$T$
region. The ratio of $F_h^{'x}(q)$ at 2T and 6T is $\sim$ 5/2,
explainable by the shifting of the Fermi point in $q$ as a
function of $B$ and the asymmetric position of the $^{13}$C probe
between two magnetic Cu$^{2+}$ ions.

To summarize, based on the conformal field theory, we presented a
comprehensive calculation of both the transverse and the
longitudinal DSF of the AFHC model in the LL regime. Considering
the hyperfine coupling as a free parameter, the comparison to NMR
results on the AFHC system CuPzN was performed for temperatures
$0.19<(k_B T)/J<4.5$ and fields up to $B \sim 2.2 J/\hbar
\gamma_e$. The spin dynamics close to the quantum critical point
can be extraordinary well described by the staggered part of the
transverse DSF. At low fields contributions from the homogeneous
part of the DSF have to be taken into account. We thank
T.~Giamarchi for suggesting us to perform this study, L.I.~Glazman
for the discussions on the range of applicability of exponents,
and W.~Brenig for the discussion on the QMC. This work was
supported by the DFG through Grant No. KL1086/8-1 of FOR 912. AAZ
acknowledges the financial support by the DFG via the Mercator
program.


\begin{thebibliography}{99}

\bibitem{NatureFocus2008} S.~Sachdev, {\it Quantum Phase Transitions}
(Cambridge University Press, Cambridge 1999); Nature Physics Focus,
{\bf 4}, 157 (2008).

\bibitem{Klanjsek} M.~Klanj\v{s}ek {\it et al.},
%H.~Mayaffre, C.~Berthier, M.~Horvati\'{c},
%B.~Chiari, O.~Piovesana, P.~Bouillot, C.~Kollath, E.~Orignac, R.~Citro, and
%T.~Giamarchi,
Phys. Rev. Lett. {\bf 101}, 137207 (2008); B.~Lake {\it et al.},
%D.A.~Tennant, C.D.~Frost, and S.E.~Nagler,
Nature Materials {\bf 4}, 329 (2005).

\bibitem{Borsa} F.~Borsa, and M.~Mali, Phys. Rev. B {\bf 9}, 2215 (1974);
M.~Takigawa {\it et al.},
%N.~Motoyama, H.~Eisaki, and S.~Uchida,
Phys. Rev. Lett. {\bf 76}, 4612 (1996); A.U.B.~Wolter {\it et al.},
%P.~Wzietek, S.~S\"{u}llow, F.J.~Litterst, A.~Honecker,
%W.~Brenig, R.~Feyerherm, and H.-H.~Klauss,
{\it ibid.} {\bf 94}, 057204 (2005).

\bibitem{book} {\it Quantum Magnetism}, ed. by U.~Schollw\"ock, J.~Richter, D.J.J.~Farnell,
and R.F.~Bishop, (Springer, Berlin-Heidelberg 2004).

\bibitem{Kuehne} H.~K\"{u}hne {\it et al.},
%H.-H.~Klauss, S.~Grossjohann, W.~Brenig,
%F.J.~Litterst, A.P.~Reyes, P.L.~Kuhns, M.M.~Turnbull, and C.P.~Landee,
Phys. Rev. B {\bf 80}, 045110 (2009); H.~K\"{u}hne {\it et al.},
%M.~G\"{u}nther, S.~Grossjohann, W.~Brenig,
%F.J.~Litterst, A.P.~Reyes, P.L.~Kuhns, M.M.~Turnbull, C.P.~Landee,
%and H.-H.~Klauss,
Physica Status Solidi {\bf B 247}, 671 (2010).

\bibitem{lancaster} T.~Lancaster {\it et al.},
%S.J.~Blundell, M.L.~Brooks, P.J.~Baker, F.L.~Pratt,
%J.L.~Manson, C.P.~Landee, and C.~Baines,
Phys. Rev. B {\bf 73}, 020410(R) (2006); P.R.~Hammar {\it et al.},
%M.B.~Stone, D.H.~Reich, C.~Broholm, P.J.~Gibson,
%M.M.~Turnbull, C.P.~Landee and M.~Oshikawa,
{\it ibid} {\bf 59}, 1008 (1999); M.B.~Stone {\it et al.},
%D.H.~Reich, C.~Broholm, K.~Lefmann, C.~Rischel,
%C.P.~Landee, and M.M.~Turnbull,
Phys. Rev. Lett. 91, 037205 (2003).


\bibitem{Moriya1956} T.~Moriya, Prog. Theor. Phys. {\bf 16}, 23 (1956).

\bibitem{MTBB} G.~M\"uller {\it et al.},
%H.~Thomas, H.~Beck, and J.C.~Bonner,
Phys. Rev. B {\bf 24}, 1429 (1981).

\bibitem{CM} J.-S.~Caux, and J.-M.~Maillet, Phys. Rev. Lett. {\bf 95}, 077201
(2005).

\bibitem{Zb} See, e.g., A.A.~Zvyagin, {\it Finite Size Effects in Correlated
Electron Models: Exact Results} (Imperial College Press, London,
2005).

\bibitem{LZ} S.~Lukyanov and A.~Zamolodchikov, Nucl. Phys. B {\bf
493}, 571 (1997); S.~Lukyanov, Phys. Rev. B {\bf 59}, 11163 (1999);
V.~Barzykin, {\it ibid.} {\bf 63}, 140412(R) (2001); T.~Hikihara and
A.~Furusaki, {\it ibid.} {\bf 69}, 064427 (2004).

\bibitem{Z} A.A.~Zvyagin, Phys. Rev. B {\bf 81}, 224407 (2010).

\bibitem{BETG} M.~Bocquet {\it et al.},
%F.H.L.~Essler, A.M.~Tsvelik, and A.O.~Gogolin,
Phys. Rev. B {\bf 64}, 094425 (2001).

\bibitem{PKKG} M.~Pustilnik {\it et al.},
%M.~Khodas, A.~Kamenev, and L.I.~Glazman,
Phys. Rev. Lett. {\bf 96}, 196405 (2006); R.G.~Pereira {\it et al.},
%J.~Sirker, J.-S.~Caux, R.~Hagemans,
%J.M.~Maillet, S.R.~White, and I.~Affleck,
{\it ibid.} {\bf 96}, 257202 (2006); M.~Khodas {\it et al.},
%M.~Pustilnik, A.~Kamenev, and L.I.~Glazman,
{\it ibid.} {\bf 99}, 110405 (2007);  R.G.~Pereira, S.R.~White,
and I.~Affleck, {\it ibid.} {\bf 100}, 027206 (2008);
A.~Imambekov, and L.I.~Glazman, Science {\bf 323}, 228 (2009);
A.~Imambekov, and L.I.~Glazman, Phys. Rev. Lett. {\bf 102}, 126405
(2009); T.L.~Schmidt, A.~Imambekov, and L.I.~Glazman, {\it ibid.}
{\bf 104}, 116403 (2010).

\bibitem{Validov2010} A.A.~Validov {\it et al.},
%E.M.~Lavrentyeva, M.~Ozerov, S.A.~Zvyagin,
%M.M.~Turnbull, C.P.~Landee, and G.B.~Teitel'baum,
J.Phys.: Conf. Series {\bf 200}, 022070 (2010).

\end{thebibliography}
\end{document}